# Tuning the Berry curvature in 2D Perovskite

Laura Polimeno[1,2], Milena De Giorgi[2], Giovanni Lerario[2], Luisa De Marco[2], Lorenzo Dominici[2], Vincenzo Ardizzone[2], Marco Pugliese[1,2], Carmela T. Prontera[2], Vincenzo Maiorano[2], Anna Moliterni[3], Cinzia Giannini[3], Vincent Olieric[6], Giuseppe Gigli[1,2], Dario Ballarini[2], Dmitry Solnyshkov[4,5], Guillaume Malpuech[4], Daniele Sanvitto[2,7]

[1]Dipartimento di Matematica e Fisica, "Ennio de Giorgi", Università Del Salento, Campus Ecotekne, via Monteroni, Lecce, 73100, Italy.

[2]CNR NANOTEC, Institute of Nanotechnology, Via Monteroni, 73100 Lecce, Italy.

[3]Istituto di Cristallografia, CNR, Via Amendola, 122/O Bari, 70126 Italy.

[4]Institut Pascal, PHOTON-N2, Université Clermont Auvergne, CNRS, SIGMA Clermont, F-63000 Clermont-Ferrand, France.

[5]Institut Universitaire de France (IUF), 75231 Paris, France.

[6]Paul Scherrer Institute, Forschungstrasse 111, Villigen-PSI, 5232, Switzerland.

[7] INFN Istituto Nazionale di Fisica Nucleare, Sezione di Lecce, 73100 Lecce, Italy.

## Abstract

Topological physics and in particular its connection with artificial gauge fields is a forefront topic in different physical systems, ranging from cold atoms to photonics and more recently semiconductor dressed exciton-photon states, called polaritons. Engineering the energy dispersion of polaritons in microcavities through nanofabrication or exploiting the intrinsic material and cavity anisotropies has demonstrated many intriguing effects related to topology and emergent gauge fields. Here, we show that we can control the Berry curvature distribution of polariton bands in a strongly coupled organic-inorganic 2D perovskite single crystal. The spatial anisotropy of the perovskite crystal combined with photonic spin-orbit coupling make emerge two Hamilton's diabolical points in the dispersion. The application of an external magnetic field breaks time reversal symmetry thanks to the exciton Zeeman splitting. It splits the diabolical points degeneracy. The resulting bands show non-zero integral Berry curvature which we directly measure by state tomography. Crucially, we show that we can control the Berry curvature distribution in the band, the so-called band geometry, within the same microcavity.

## I. INTRODUCTION

The generation of artificial gauge fields (AGF) is a central topic in modern physics and has been recently investigated in a great variety of physical platforms [1], such as ultra-cold atomic gases [2, 3, 4], photonic crystals [5, 6, 7], graphene and graphene like materials [8, 9], mechanical systems [10] and exciton-polaritons [11, 12, 13]. One of the main concepts behind the theory of AGF in photonic systems lies in the topological and geometrical properties of bands associated to the buildup of a non-zero Berry curvature. The Berry curvature can be considered as a pseudomagnetic field in the reciprocal space and it is determined by the wave function change along particle energy dispersion. It is included in a more general object, the quantum geometric tensor (QGT), whose symmetric real part defines a metric and gives information about the distance between the eigenstates, while its antisymmetric imaginary part represents the Berry curvature [14, 15].

The Berry curvature, whose integral is a topological invariant called the Chern number, plays a key role in topological physics, while the quantum metric is important for understanding many condensed matter phenomena such as superfluidity in flat bands [16], orbital magnetic susceptibility [17], the exciton Lamb shift [18] and the non-adiabatic anomalous Hall effect [19, 20, 21]. To achieve a non-zero integral value of the Berry curvature for a band in a photonic system, both the optical spin-orbit coupling and the breakdown of the time-reversal symmetry (TRS) are necessary [21]. The combination of such geometrically non-trivial bands and of a band gap then allows the opening of a topological gap [22, 23].

Recently, the frontiers of this field have been extended to the exciton-polariton systems, exploiting their unique properties as part-light, part-matter quasi-particles (polaritons), resulting from the strong coupling of excitons and photons. The importance of the exciton-polaritons [23] lies in the possibility to have a high degree of freedom in the engineering of the particles's Hamiltonian when combining the physical phenomena associated to the exciton component, as the exciton Zeeman effect, with those due to the photon energy dispersion in optical confined systems, in particular, photonic spin-orbit coupling [24]. Several strategies to realize topological systems with polaritons have been suggested [25, 26, 27], by using artificial lattices with honeycomb geometries that have been shown to support Dirac cone dispersions [28, 29], as well as edge modes [30], inherited from their graphene-like structural origin. Recently, artificial lattices have been realized both in organic [31] and perovskite materials [32], paving the way for the realization and control of quantum states stable at room temperature, thanks to the intrinsic robustness of the exciton in these materials.

In this context, polariton systems based on 2D hybrid organic-inorganic perovskites [33, 34, 35] represent a unique platform for topological studies, thanks to the possibility to easily tune the optoelectronic properties of the polariton device through substitution of the organic interlayer [36], without the need of complex fabrication techniques as in previous works, such as electron beam lithography and dry etching. Moreover, they have an exciton with large binding energies, which make these materials suitable to easily work at room temperature and display stronger dielectric confinement compared to inorganic materials.

In this work, we study the topological properties of an exciton-polariton planar resonator based on an optical birefringent 2D perovskite. In particular, we demonstrate that a non-zero integral value of the Berry curvature is obtained in presence of an external magnetic field, necessary to break the TRS, making use of the exciton Zeeman splitting. Furthermore, we perform direct measurements of the Berry curvature distribution, the so-called band geometry. We demonstrate that the band geometry can be controlled through the exciton/photon fractions of the polariton mode, the quantities which we can widely tune taking advantage of the limited energy spacing between the confined optical modes of our microcavity.

Our findings pave the way for the formation of topological states and manipulation of their topological properties in perovskite-based polariton systems.

## II. RESULTS AND DISCUSSION

The 4-fluorophenethylammonium lead iodide ($F-(C_6H_5(CH_2)_2NH_3)_2PbI_4$) perovskite crystals (PEAI-F) have been synthesized by anti-solvent vapor assisted crystallization method [37,36] on the top of a Distributed Bragg Reflector (DBR) made by seven $SiO_2/TiO_2$ pairs. Single-crystal X-ray diffraction data measurements performed on PEAI-F crystals (see the crystal packing in Fig. S1, the asymmetric unit in Fig. S2A and X-Ray Diffraction Section in Supporting Information for further details) highlight the in-plane distortion of the inorganic layers with octahedra tilting (see Fig. S1A, S2B and Table S2 of the Supporting Information), which could favour the optical birefringence of the material. After mechanical exfoliation performed in order to obtain crystals with thickness between 3 $\mu m$ and 7 $\mu m$, the planar cavity has been closed by evaporating an 80 *nm*-thick silver layer as top mirror (Fig. 1A).

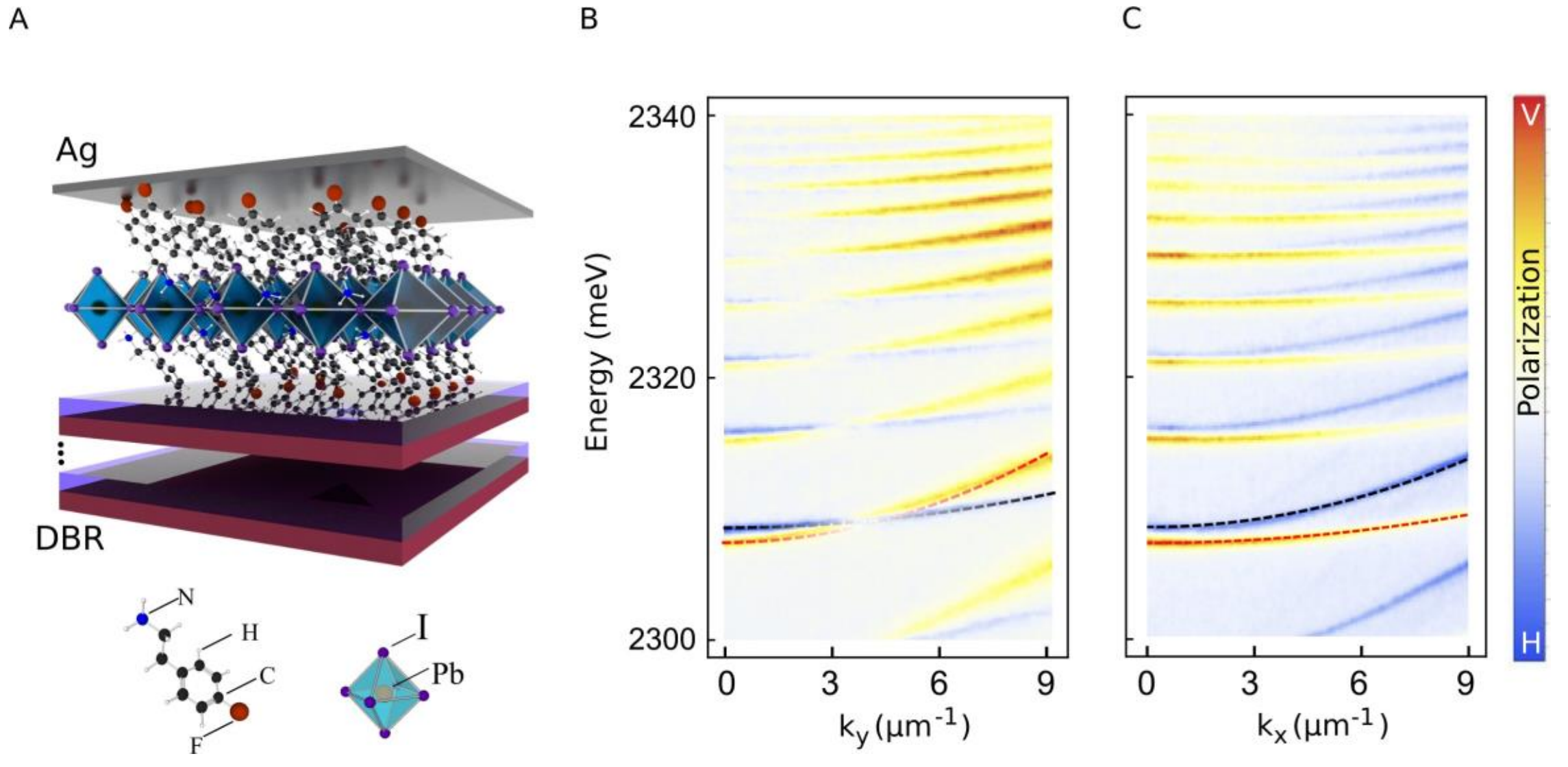

Figure 1: A) Schematic representation of the sample. Perovskite crystals are embedded in a planar microcavity made by a bottom DBR with seven $TiO_2/SiO_2$ pairs and a top 80 *nm* thick silver mirror. B, C) Degree of polarization (H/V) of the photoluminescence signal of the lower polariton branches resolved in the energy vs $k_y$ (B) and $k_x$ (C) in-plane momentum space. The perovskite thickness is 7 μm. The dashed lines are the fitting of the system eigenstates derived from eq.2. with zero magnetic field ($\Delta_z$=0).

The sample was inserted in a cryostat, cooled down to liquid helium for high magnetic field measurements. The sample is non-resonantly excited with a continuous wave laser at $\lambda$ = 488 *nm* (2541 *meV*). Polarized resolved photoluminescence (PL) was measured at 0 T and 9 T. Figs. 1B,C show the energy dispersion versus the in-plane momentum, obtained from the photoluminescence spectra as $(I_H - I_V)/(I_H + I_V)$ along the $k_y$ (Fig. 1B) and $k_x$ (Fig. 1C) directions at 0 *T*, where $I_H$ and $I_V$ are the PL intensities of the horizontal and vertical polarizations, respectively. A manifold of modes, which come in pairs due to a linear polarization splitting, results from the free spectral range of the microcavity, filled with a 7 $\mu m$ thick perovskite layer (Figs. 1B, C). Neglecting polarization splitting, the energy of the bare photon modes of the Fabry-Perot resonator at k=0 is given by $E_j \approx j\frac{\hbar c\pi}{Ln_r}$ where $j$ is the mode number, $L$ the effective cavity thickness, $n_r$ the effective refractive index of the cavity. Here $j \sim 200$ and the flattening of the mode dispersion that we observe while approaching the exciton resonance at 2.382 $eV$ (see Fig. S3), is due to the strong coupling between the bare exciton of the PEAI-F single crystal and the cavity modes, and not to the weak increase of the photon mode effective mass which is proportional to $j$ and which is less than 5% on the displayed energy scale. Therefore, the differences in the polariton effective mass between different modes is caused by the increase of the exciton fraction. The energy dispersion of a given exciton-polariton mode can be approximately found by diagonalizing a two by two Hamiltonian, describing two coupled linear oscillators. The energy distance between the polariton modes decreases approaching the exciton energy, because it comes from the splitting of the Fabry Perot modes and their photon fraction $p_j$ decreases. In the same way the polariton's effective mass is $\sim m_j^p / p_j$ where $m_j^p$ is the bare photon mass of the mode $j$. We now concentrate on a given polarization doublet.

The different energy-momentum dispersion of the two modes in a given pair is related to the TE and TM (Transverse Electric, Transverse Magnetic) polarization anisotropy, resulting in an energy splitting of the polariton modes in the order of few meV, which increases at higher wavevector k. Such momentum dependent splitting is due both to the intrinsic asymmetry of the cavity and to the difference between the in- and out-of-plane refractive indexes of the perovskite [38]. In the absence of any other refractive index asymmetry, the splitting should be zero at zero momentum $\{k_x, k_y\} = \{0, 0\}\ \mu m^{-1}$ leading to an energy degeneracy of the TE-TM modes at this momentum space coordinate. In the PEAI-F-based sample, the presence of the fluorine at the termination of the interlayer in between the perovskite wells introduces an in-plane optical anisotropy, leading to an additional energy splitting (X-Y splitting) at the $\{k_x, k_y\} = \{0, 0\}\ \mu m^{-1}$ point, shifting the degeneracy points (Hamilton's diabolical points) towards higher wavevectors (Fig. 1B) in one in-plane k direction ($k_y$), while no crossing points are visible in the orthogonal direction ($k_x$) (Fig. 1C). In other terms, this demonstrates that in our perovskite-based system the cylindrical symmetry of the polariton energy bands is broken [38]. The X-Y energy splitting for the lowest observed energy polariton mode is about 2 *meV*, and this value decreases when the photon fraction decreases (see Fig. S4B), because it is due to the interplay of the difference of the background refractive indices. The extracted values of the X-Y energy splitting in the observed energy range in such perovskite system are about two orders of magnitude higher than the typical energy splitting observed in inorganic GaAs based microcavities (15 – 30 $\mu eV$) [21, 39].

The dashed blue and red lines in Figs. 1B, C represent the energy eigenvalues of the system for a pair of orthogonally polarized polariton modes as computed by diagonalizing an effective two band Hamiltonian accounting for both the TE-TM and X-Y anisotropy with the addition of a Zeeman splitting opened by the presence of an external magnetic field acting on the exciton component [40] [41]:

$$H_K = \begin{pmatrix} \frac{\hbar^2 k^2}{2m} + \Delta_z & \alpha - \beta k^2 e^{-2i\varphi} \\ \alpha - \beta k^2 e^{-2i\varphi} & \frac{\hbar^2 k^2}{2m} - \Delta_z \end{pmatrix} \tag{1}$$

where m is the polariton mass, k = |k| = $\sqrt{k_x^2 + k_y^2}$ is the in-plane wavevector module, $\Delta_z$ is the polariton Zeeman splitting, α and β are the strengths of the X-Y and TE-TM splitting, respectively.

Taking also into account the bottom energy of the lower polariton mode, the system's eigenvalues are extracted by diagonalizing the Hamiltonian in (1):

$$E_\pm = E_0 + \frac{\hbar^2 k^2}{2m} \pm \sqrt{\beta^2 k^4 - \beta \alpha k^2 \cos(2\theta_k) + \frac{\alpha^2}{4} + \Delta_z^2} \tag{2}$$

where $\theta_k$ is the angle of the in-plane momentum direction, k = {k cos($\theta_k$), k sin($\theta_k$)}. The fitting parameters when no external magnetic field is applied, i.e., $\Delta_z = 0$, for the pair of modes fitted in Figs. 1 b,c are $E_0$ = 2.308 *eV*, m = 4.4 · $10^{-4}$ $m_e$, β = 3.3 · $10^{-5}$ *eV* $\mu m^2$ and α = 1.1 *meV*. By comparing the dispersion spectra along the two directions, it can also be observed the presence at $k_y = \pm 3.7\ \mu m^{-1}$ of two Hamilton's diabolical points [38] (Fig. 1b). For these states, degenerate in energy, the TE-TM energy splitting exactly compensates the X-Y splitting.

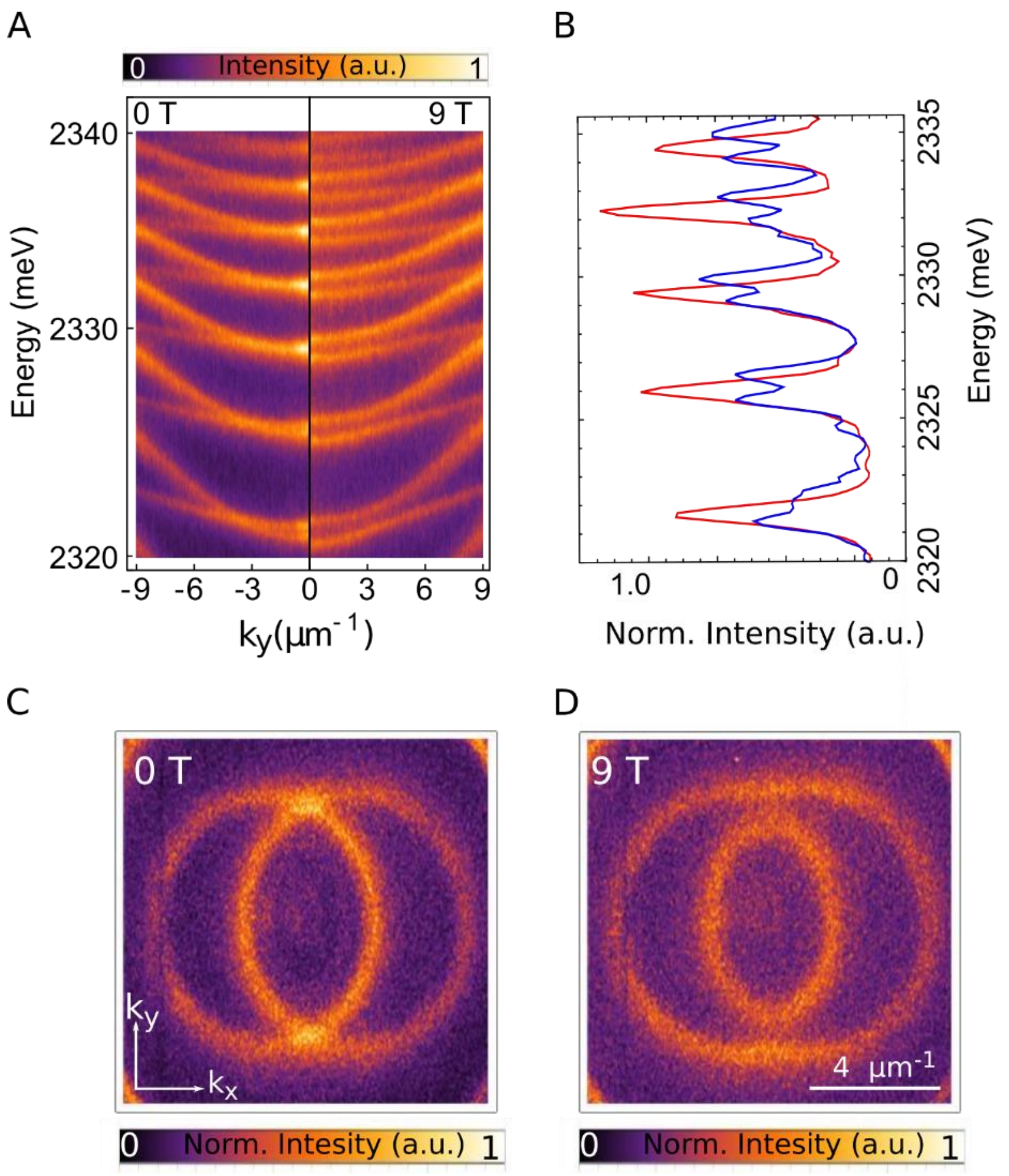

Figure 2: A) Energy vs $k_y$ in-plane momentum of unpolarized photoluminescence for $B = 0\ T$ magnetic field (right half panel) and under a magnetic field, $B = 9\ T$ (left half panel). B) Photoluminescence spectra measured at $k_x = 0\ \mu m^{-1}$ and $k_y = \pm\ 3.7\ \mu m^{-1}$ at zero magnetic field (red line) and at $B = 9\ T$ (blue line). C, D) Unpolarized photoluminescence maps in the momentum space at isoenergetic cross-sections of the lower polariton dispersion. The cuts are taken at the energy $E = 2.316\ eV$ and for a magnetic field of $B = 0\ T$ (C) and $B = 9\ T$ (D), respectively. The Zeeman effect emerging under 9T is responsible for the degeneracy suppression.

It is now possible to lift the dispersion degeneracy at the two diabolical points, by applying a magnetic field, B, perpendicular to the planar microcavity. Indeed, we can make use of a peculiar property of polaritons, their excitonic component, and remove the degeneracy through the Zeeman effect. This effect breaks the time-reversal symmetry and the polariton modes acquire a circular polarization component. By applying 9T, the spectrum opens up at the two crossing points of the polariton dispersions, generating an anti-crossing gap (Figs. 2C, D). Here, we can extract the Zeeman strength for each couple of orthogonal modes from the photoluminescence profile of the unpolarized energy dispersion spectra (Fig. 2B). We found that the Zeeman splitting increases at higher energies due to the higher exciton fraction of the polaritonic modes (see Fig. S4A), in contrast to the TE-TM and X-Y splitting, which instead decrease moving towards the exciton energy (Fig. S4B).

In order to verify that these nontrivial dispersions are consistent with the conical diffraction theory, we have measured the polarization resolved photoluminescence spectra of the modes along all the six polarization axes of the Poincaré sphere, corresponding to the horizontal-vertical (H−V), diagonal-antidiagonal (D−A), and circular right-left (R−L) polarization. In particular, at the energy $E = 2.313\ eV$, for which the two modes cross each other at the diabolical points, we observe that the linear polarization direction precesses when moving along the two crossing rings (dashed lines in Figs. S5A,B), in agreement with the conical diffraction theory. At the diabolical points, the polarization value is the linear superposition of the polarizations of the crossing modes. Along the $k_x$ direction, diametrically opposed points possess orthogonal polarization for each ring and the polarization map is centrosymmetric with respect to the origin of the momentum plane $\{k_x, k_y\} = \{0, 0\}\ \mu m^{-1}$.

The full knowledge of the mode polarization is a fundamental parameter to have an insight on the AGF theories in polariton microcavities since it defines the particle pseudospin (Stokes vector). This can be thought as a two-degree of freedom phase associated to the particles, or in other terms as a vector charge [38] property. The definition of an AGF directly derives from the Hamiltonian in (1), which can be expressed in terms of a kinetic term and an additional term associated to an effective magnetic field acting on the particle pseudospin [21]. As a further consequence, the full knowledge of the pseudospin makes possible to extract the quantum geometric tensor (QGT) [40]. Such tensor contains the structural information about the bands of the system and, in particular, how the orientation of the pseudospin changes when moving along the momentum plane. The orientation of the Stokes vector $\mathbf{S}(\mathbf{k})$, which represents the polarization state as a point on the Poincaré sphere, is determined by the equations:

$$S_1(k) = \frac{I_H - I_V}{I_H + I_V} \quad S_2(k) = \frac{I_D - I_A}{I_D + I_A} \quad S_3(k) = \frac{I_R - I_L}{I_R + I_L} \tag{3}$$

where $I_H$ and $I_V$ are the emission intensity for the horizontal and vertical polarization, $I_D$ and $I_A$ for the diagonal and antidiagonal polarization and finally, $I_R$ and $I_L$ are the intensity of the right and left circular polarization, respectively. It is hence needed to retrieve these polarization-resolved intensities in the full $\{k_x, k_y, E\}$ domain, in order to extract the pseudospin **S** components for a given energy band at each wavevector **k**. The unitary vector can also be written in terms of the polar angle $\vartheta(\mathbf{k}) = \arccos S_3(\mathbf{k})$ and the azimuthal angle $\varphi(\mathbf{k}) = \arctan[S_2(\mathbf{k})/S_1(\mathbf{k})]$ on the Poincaré sphere, which allows to reconstruct the components of the QGT, $g_{ij}$, and the Berry curvature, $B_z$, in the momentum space:

$$g_{ij} = \frac{1}{4}\left(\partial_{k_i}\theta\,\partial_{k_j}\theta + \sin^2\theta\;\partial_{k_i}\varphi\,\partial_{k_j}\varphi\right) \tag{4}$$

$$B_z = \sin\theta\;\frac{1}{2}\left(\partial_{k_i}\theta\,\partial_{k_j}\varphi - \partial_{k_j}\theta\,\partial_{k_i}\varphi\right) \tag{5}$$

which are both related to the pseudospin texture in the momentum space. In particular, the quantum metric tensor can be used in the definition of a distance between the eigenstates.
Note that using the effective Hamiltonian (1), QGT components can be computed analytically [40], in particular, the Berry curvature of two polarization bands reads:

$$B^{j}_{z\pm} = \frac{\pm 2\,\beta\,k^2\,\Delta_z}{\left(\alpha^2+2\left(k_y^2-k_x^2\right)\alpha\beta+k^4\beta^2\right)+\Delta_z^2} \tag{6}$$

This formula shows that the tuning of the parameters α, β and $\Delta_z$, given by the different exciton/photon fraction at each mode, allows a remarkable control of the Berry curvature distribution in reciprocal space, that is, of the so-called band geometry, as we demonstrate experimentally below.

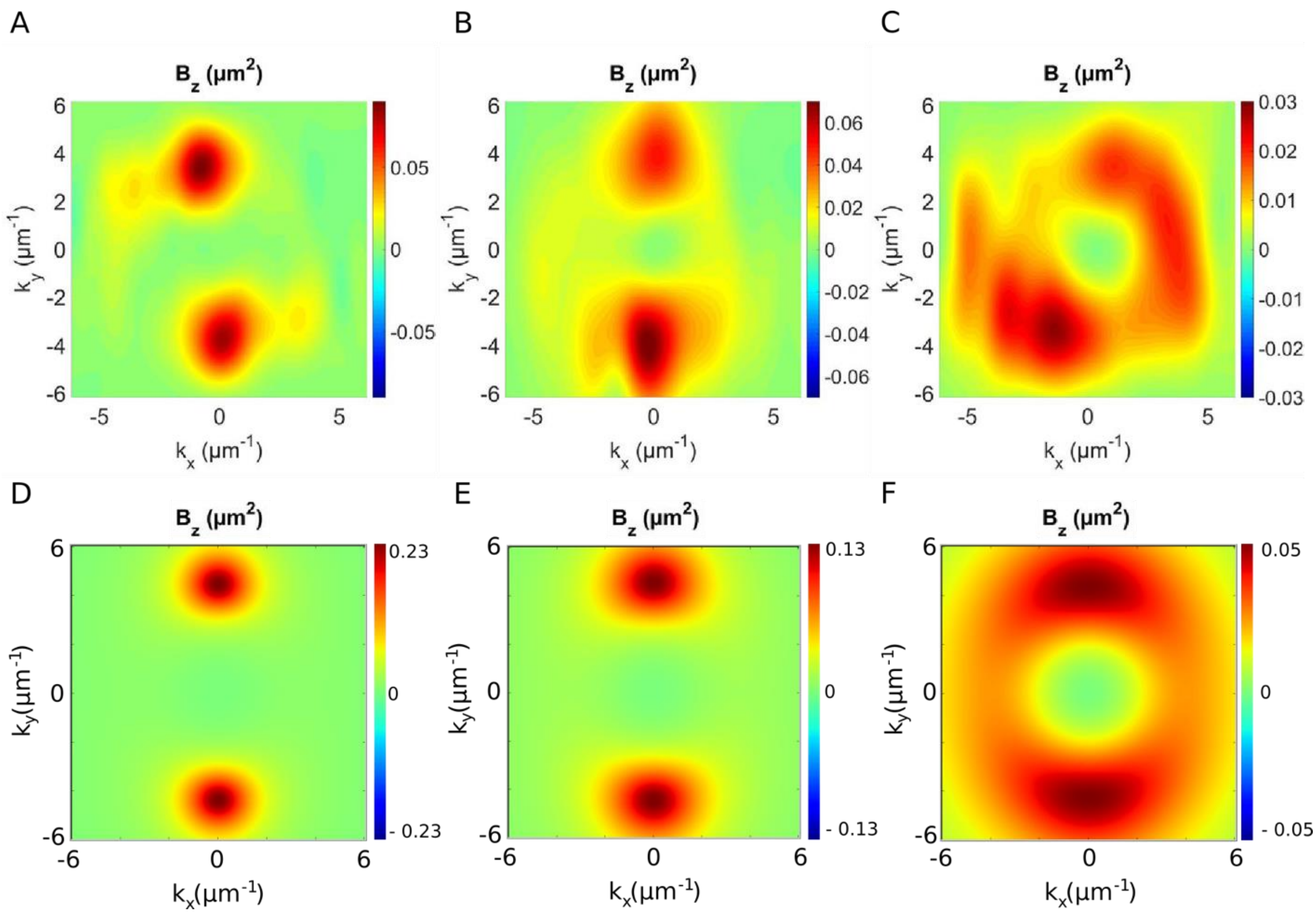

Figure 3: A-F) Experimental and theoretical Berry curvature extracted from the polarization-resolved measurements, for mode j (A, D), mode j+1 (B, E) and mode j+2 (C, F), respectively.

Applying 9T magnetic field, the time reversal symmetry is broken and polarization bands anti-cross instead of crossing. The Berry curvature which was concentrated at the Hamilton crossing points gets distributed in the two split bands (with a given sign for a given branch) and in k-space. The precise distribution realized depends on the polariton exciton and photon fractions.

Figs. 3A-C show the experimental Berry curvature distribution from the 3 modes j, j+1, j+2 whose dispersion is shown in the Fig. S6 of the Supporting Information. The Berry curvature extraction is performed as indicated above, by measuring the polarization resolved emission of each state of the 2D dispersion which allows a full determination of the pseudo-spin of the eigenstates. Then formula 5 is applied.

In Fig. 3A, the Berry curvature is concentrated around the two anti-crossing points, forming two broadened Berry monopoles. The shape evolves, becoming ring like while going to more exciton like polaritons. As one can visualize in Figs. 3 D-F, this measured behavior is in good agreement with the simulated Berry curvature, considering the experimental values of α, β and $\Delta_z$, extracted for each polariton mode. The evolution of the Berry curvature can be understood taking into account that at higher energies the exciton fraction of the polariton modes increases, resulting in a more pronounced Zeeman splitting [40] and in a decrease of the TE-TM and X-Y splitting effects due to the dispersion flattening at the exciton transition. It is the opposite trend of the contributions to the effective magnetic field with respect to the photon-exciton fraction of the polariton modes, which makes ultimately possible to observe Berry curvatures with different shapes inside the same device. It is interesting to note that for mode j+1 it is possible to observe a localized Berry curvature even for zero magnetic field (see Fig. S5C), probably resulting from a weak optical chirality of the perovskite layer, similarly to [42].

## III. CONCLUSIONS

We have demonstrated the possibility to tune the topological properties of a polariton system through the use of different band geometries in a perovskite microcavity. By combining the material's optical birefringence with the Zeeman effect, we observed the switching of the Berry curvature in momentum space from two well-defined maxima, to a cylindrically symmetric distribution. Such kind of structures can be used for multiplexing the future optical valleytronic devices, such as transistors based on the anomalous Hall effect, operating at different wavelengths at the same time.

## IV. METHODS

**Synthesis of** 2*D* **perovskite flakes** A solution 0.5 M PEAI-F is prepared in a nitrogen- filled glovebox by dissolving $PbI_2$ and 4-Fluorophenethylammonium iodide (1:2 molar ratio) in *γ*-butyrolactone and stirring at 70°*C* for 1 hour. 2D perovskite single crystals are synthesized using an anti-solvent vapor-assisted crystallization method as follows: 3 $\mu$l of perovskite solution are deposited on top a sputtered DBR and covered with a glass coverslip. Substrates and a small vial containing 2 ml of dichlorometane are placed inside a bigger Teflon vial which is closed with a screw cap and left undisturbed for 12 hours at room temperature. During this time, crystals slowly grow in a saturated environment of dichloromethane (antisolvent) and at the end millimeter-sized perovskite flakes appear on the top of the DBR. Their thickness varies from few to tens of micrometers. Using SPV 224PR-M Nitto Tape or PDMS, mechanical exfoliation is carried out on the perovskite flakes in order to obtain single crystals having the desired thickness.

**Microcavity Sample Fabrication** The DBR is made by seven pairs of $TiO_2/SiO_2$ (63 *nm/*94 *nm*) deposited by radio-frequency (RF) sputtering process— in an Argon atmosphere under a total pressure of $6 \cdot 10^3$ mbar and at RF power of 250 *W* —on top of a 1 *mm* glass substrate. The perovskite single crystals are grown on top of the DBR (see above) and an 80 nm-thick layer of silver is thermally evaporated on top of the structure (deposition parameters: current = 280 *A*, deposition-rate = 3 Å/*s*).

**X-ray diffraction** Single-crystal X-ray diffraction data measurements for PEAI-F were carried out at the beamline PXIII (X06DA-PXIII, http://www.psi.ch/sls/pxiii/) at the Swiss Light Source (SLS), Villigen, Switzerland, using a Parallel Robotics Inspired (PRIGo) multiaxis goniometer [43] and a PILATUS 2M-F detector. Data collection was performed at low temperature (T = 100 K) on a selected crystal mounted on lithoIoops (Molecular Dimensions). Complete data were obtained by merging two 360° $\omega$ scans at $\chi$=0° and $\chi$=30° of PRIGo. In shutterless mode, a 360° data set was collected in 3 min (beam energy of 17 keV, $\lambda$=0.72932 Å, focus size 90 x 50 $\mu m^2$, 0.25 sec of exposure time per frame, 0.5° scan angle). Main experimental details are given in Table S1. Diffraction data were processed by *XDS* [44], a software organized in eight subroutines able to perform the main data reduction steps; the integrated intensities were scaled and corrected for absorption effects by the *XSCALE* subroutine [44].

Structure solution was carried out by Direct Methods [45] using *SIR2019* [46] a package that exploits the information on unit cell parameters, diffraction intensities and expected chemical formula to identify the space group and determine the crystal structure by Direct Methods. The partial structure model located by *SIR2019* was completed and refined using full-matrix least-squares techniques by *SHELXL2014/7* [47]. Non-hydrogen atoms were refined anisotropically. Hydrogen atoms were positioned by difference Fourier map; their atomic coordinates were freely refined and the following constraints on the isotropic U value of H atoms were applied: $U_{iso}$(H)=1.2 $U_{eq}$(C) and $U_{iso}$(H)=1.5 $U_{eq}$(N) in case of hydrogen atoms bonded to C and N atoms, respectively.

**Optical setup** Figure 4 shows a sketch of the optical setup. The measurements are performed in reflection configuration at cryogenic temperature (T=4K) using a continuous- wave laser at 488 *nm* and an optical chopper (with rotation frequency = 300 *Hz*), to minimize the sample heating.

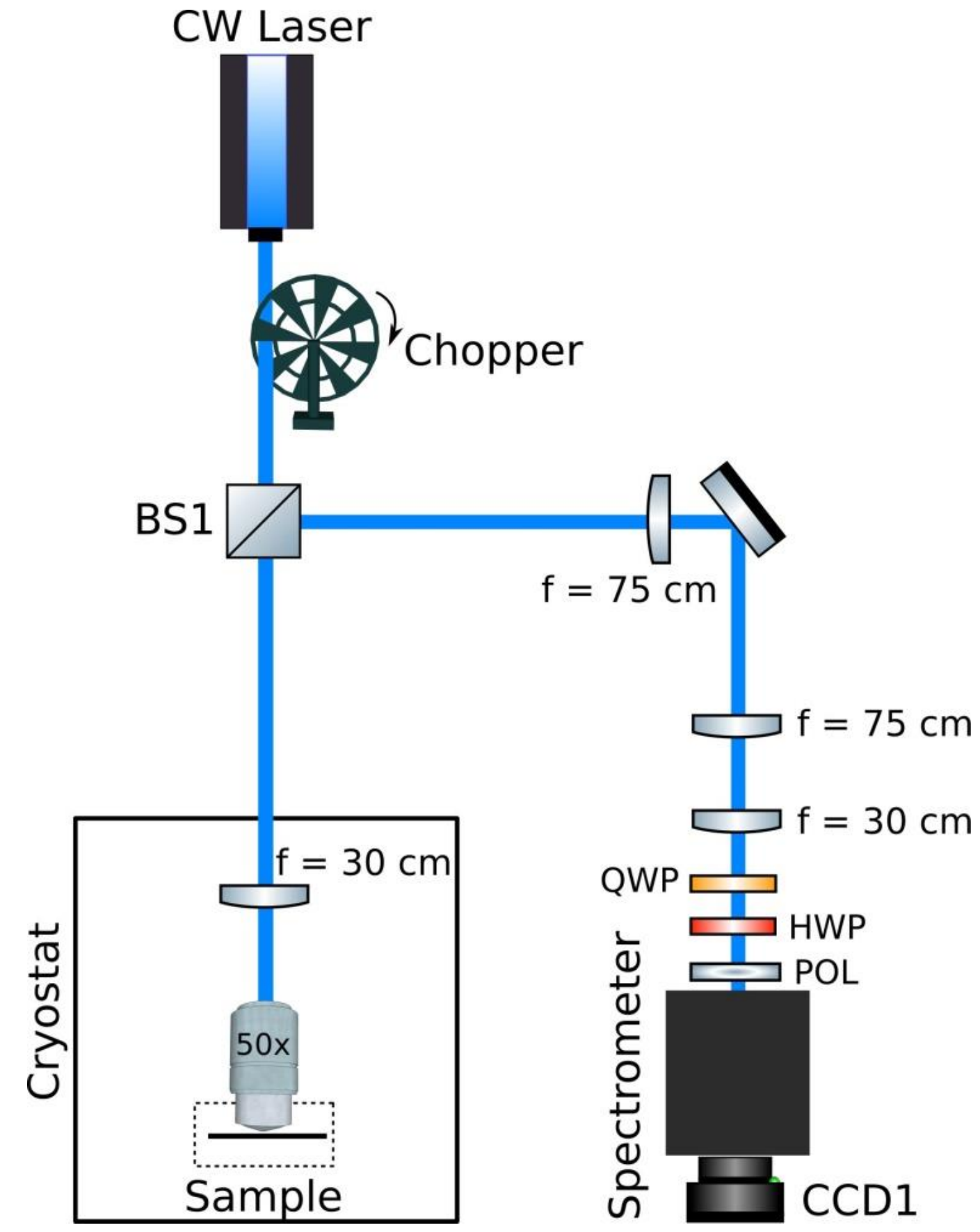

Figure 4: Sketch of the optical setup

Through the detection path, the image on the back focal plane of the detection objective (50 *x* with *N.A.* = 0.82) is projected on a spectrometer entrance slit. The spectrometer is coupled to an enhanced CCD camera for the detection of the polariton energy dispersion. The second 30*cm* lens is on a translation stage in order to scan the whole 2*D*-momentum space. The quarter-wave plate (QWP) and the half-wave plate (HWP) in front of the entrance of the spectrometer allow for the detection of the polarization-resolved maps in momentum space.

## Acknowledgments

The authors acknowledge Paolo Cazzato for technical support, Iolena Tarantini for the metal evaporation and Dario Gerace for useful discussions. The authors acknowledge the project PRIN Interacting Photons in Polariton Circuits – INPhoPOL (Ministry of University and Scientific Research (MIUR), 2017P9FJBS_001) and the project "TECNOMED - Tecnopolo di Nanotecnologia e Fotonica per la Medicina di Precisione", (Ministry of University and Scientific Research (MIUR) Decreto Direttoriale n. 3449 del 4/12/2017, CUP *B*83*B*17000010001). G.G. gratefully acknowledges the project PERSEO- PERrovskite-based Solar cells: towards high Efficiency and lOng-term stability (Bando PRIN 2015-Italian Ministry of University and Scientific Research (MIUR) Decreto Direttoriale 4 novembre 2015*n.*2488, project number 20155*LECAJ* ). We also acknowledge the support of the project "Quantum Fluids of Light" (ANR-16-CE30-0021), of the ANR Labex GaNEXT (ANR-11-LABX-0014), and of the ANR program "Investissements d'Avenir" through the IDEX-ISITE initiative 16-IDEX-0001 (CAP 20-25).

---

[1] M. Aidelsburger, S. Nascimbene, and N. Goldman, "Artificial gauge fields in materials and engineered systems," *Comptes Rendus Physique*, vol. 19, no. 6, 2018.

[2] J. Dalibard, "Introduction to the physics of artificial gauge fields", *arXiv*:1504.05520 [cond-mat.quant-gas], 2015.

[3] N. Goldman, J. C. Budich, and P. Zoller, "Topological quantum matter with ultracold gases in optical lattices," *Nature Physics*, vol. 12, no. 7, 2016.

[4] N. Goldman, G. Juzeliunas, P. Öhberg, and I. B. Spielman, "Light-induced gauge fields for ultracold atoms," *Reports on Progress in Physics*, vol. 77, no. 12, 2014.

[5] L. Lu, J. D. Joannopoulos, and M. Soljacic, "Topological photonics," *Nature Photonics*, vol. 8, no. 11, 2014.

[6] M. Hafezi, "Synthetic gauge fields with photons," *International Journal of Modern Physics B*, vol. 28, no. 02, 2014.

[7] T. Ozawa, H. M. Price, A. Amo, N. Goldman, M. Hafezi, L. Lu, M. C. Rechtsman, D. Schuster, J. Simon, O. Zilberberg *et al.*, "Topological photonics," *Reviews of Modern Physics*, vol. 91, no. 1, 2019.

[8] M. A. Vozmediano, M. Katsnelson, and F. Guinea, "Gauge fields in graphene," *Physics Reports*, vol. 496, no. 4-5, 2010.

[9] Y. Ren, Z. Qiao, and Q. Niu, "Topological phases in two-dimensional materials: a review," *Reports on Progress in Physics*, vol. 79, no. 6, 2016.

[10] S. D. Huber, "Topological mechanics," *Nature Physics*, vol. 12, no. 7, 2016.

[11] T. Gao, E. Estrecho, K. Bliokh, T. Liew, M. Fraser, S. Brodbeck, M. Kamp, C. Schneider, S. Höfling, Y. Yamamoto *et al.*, "Observation of non-hermitian degeneracies in a chaotic exciton-polariton billiard," *Nature*, vol. 526, no. 7574, 2015.

[12] E. Estrecho, T. Gao, S. Brodbeck, M. Kamp, C. Schneider, S. Höfling, A. Truscott, and E. Ostrovskaya, "Visualising berry phase and diabolical points in a quantum exciton-polariton billiard," *Scientific reports*, vol. 6, no. 37653, 2016.

[13] H.-T. Lim, E. Togan, M. Kroner, J. Miguel-Sanchez, and A. Imamoglu, "Electrically tunable artificial gauge potential for polaritons," *Nature communications*, vol. 8, no. 14540, 2017.

[14] M. Berry, "The quantum phase five years after", *Geometric Phases in Physics*, World Scientific, 1989.

[15] M. Berry, "Quantal phase factors accompanying adiabatic changes", *Proceedings of the Royal Society of London. A. Mathematical and Physical Sciences*, vol. 392, no. 1802, 1984.

[16] S. Peotta and P. Törmä, "Superfluidity in topologically nontrivial flat bands", *Nature Communications*, vol. 6, no. 8944, 2015.

[17] F. Piéchon, A. Raoux, J. N. Fuchs, and G. Montambaux, "Geometric orbital susceptibility: Quantum metric without Berry curvature", *Physical Review B*, vol. 94, no. 134423, 2016.

[18] A. Srivastava and A. Imamoglu, "Signatures of Bloch-Band Geometry on Excitons: Nonhydrogenic Spectra in Transition-Metal Dichalcogenides", *Physical Review Letters*, vol. 115, no. 166802, 2015.

[19] N. Nagaosa, J. Sinova, S. Onoda, A. H. MacDonald, and N. P. Ong, "Anomalous Hall effect", *Reviews of Modern Physics*, vol. 82, no. 1539, 2010.

[20] J. Sinova, S. O. Valenzuela, J. Wunderlich, C. H. Back, and T. Jungwirth, "Spin Hall effects", *Reviews of Modern Physics*, vol. 87, no. 1213, 2015.

[21] A. Gianfrate, O. Bleu, L. Dominici, V. Ardizzone, M. De Giorgi, D. Ballarini, G. Lerario, K. W. West, L. N. Pfeiffer, D. D. Solnyshkov, D. Sanvitto, and G. Malpuech, "Measurement of the quantum geometric tensor and of the anomalous Hall drift", *Nature*, vol. 578, no. 7795, 2020.

[22] S. Raghu and F. D. Haldane, "Analogs of quantum-Hall-effect edge states in photonic crystals", *Physical Review A - Atomic, Molecular, and Optical Physics*, vol. 78, no. 033834, 2008.

[23] T. Karzig, C. E. Bardyn, N. H. Lindner, and G. Refael, "Topological polaritons", *Physical Review X*, vol. 5, no. 031001, 2015.

[24] A. Kavokin, G. Malpuech, and M. Glazov, "Optical spin hall effect", *Physical Review Letters*, vol. 95, no. 136601, 2005.

[25] C. E. Bardyn, T. Karzig, G. Refael, and T. C. Liew, "Topological polaritons and excitons in garden-variety systems", *Physical Review B - Condensed Matter and Materials Physics*, vol. 91, no. 161413, 2015.

[26] A. V. Nalitov, D. D. Solnyshkov, and G. Malpuech, "Polariton Z topological insulator", *Physical Review Letters*, vol. 114, no. 116401, 2015.

[27] S. Klembt, T. H. Harder, O. A. Egorov, K. Winkler, R. Ge, M. A. Bandres, M. Emmerling, L. Worschech, T. C. Liew, M. Segev, C. Schneider, and S. Höfling, "Exciton-polariton topological insulator", *Nature*, vol. 562, no. 552, 2018.

[28] T. Jacqmin, I. Carusotto, I. Sagnes, M. Abbarchi, D. D. Solnyshkov, G. Malpuech, E. Galopin, A. Lemaître, J. Bloch, and A. Amo, "Direct observation of Dirac cones and a flat band in a honeycomb lattice for polaritons", *Physical Review Letters*, vol. 112, no. 116402, 2014.

[29] B. Real, O. Jamadi, M. Milicévic, N. Pernet, P. St-Jean, G. Ozawa, I. Montambaux, I. Sagnes, A. Lemaître, L. Le Gratiet, A. Harouri, S. Ravets, J. Bloch, and A. Amo, "Semi-Dirac transport and anisotropic localization in polariton honeycomb lattices", *arXiv*: 2004.03478 *[cond-mat.mes-hall]* , 2020.

[30] M. Milicévic, T. Ozawa, G. Montambaux, I. Carusotto, E. Galopin, A. Lemaître, L. Le Gratiet, I. Sagnes, J. Bloch, and A. Amo, “Orbital Edge States in a Photonic Honeycomb Lattice”, *Physical Review Letters*, vol. 118, no. 107403, 2017.

[31] F. Scafirimuto, D. Urbonas, U. Scherf, R. F. Mahrt, and T. Stöferle, “Room-Temperature Exciton-Polariton Condensation in a Tunable Zero-Dimensional Microcavity”, *ACS Photonics*, vol. 5, no. 1, 2018.

[32] R. Su, S. Ghosh, J. Wang, S. Liu, C. Diederichs, T. C. Liew, and Q. Xiong, “Observation of exciton polariton condensation in a perovskite lattice at room temperature”, *Nature Physics,* vol. 16, no. 3, 2020.

[33] L. Pedesseau, D. Sapori, B. Traore, R. Robles, H. H. Fang, M. A. Loi, H. Tsai, W. Nie, J. C. Blancon, A. Neukirch, S. Tretiak, A. D. Mohite, C. Katan, J. Even, and M. Kepenekian, “Advances and Promises of Layered Halide Hybrid Perovskite Semiconductors”, *ACS Nano,* vol. 10, no. 11, 2016.

[34] B. Saparov and D. B. Mitzi, “Organic-Inorganic Perovskites: Structural Versatility for Functional Materials Design”, *Chem. Rev.*, vol. 10, no. 11, 2016.

[35] F. Thouin, S. Neutzner, D. Cortecchia, V. A. Dragomir, C. Soci, T. Salim, Y. M. Lam, R. Leonelli, A. Petrozza, A. R. S. Kandada, and C. Silva, “Stable biexcitons in two-dimensional metal-halide perovskites with strong dynamic lattice disorder”, *Physical Review Materials*, 2018.

[36] A. Fieramosca, L. De Marco, M. Passoni, L. Polimeno, A. Rizzo, B. L. Rosa, G. Cruciani, L. Dominici, M. De Giorgi and G. Gigli, “Tunable out-of-plane excitons in 2d single-crystal perovskites”, *ACS Photonics*, vol. 5, no. 10, 2018.

[37] F. Lédée, G. Trippé-Allard, H. Diab, P. Audebert, D. Garrot, J.-S. Lauret, and E. Deleporte, “Fast growth of monocrystalline thin films of 2d layered hybrid perovskite”, *Cryst. Eng. Comm.*, vol. 19, no. 19, 2017.

[38] A. Fieramosca, L. Polimeno, G. Lerario, L. De Marco, M. De Giorgi, D. Ballarini, L. Dominici, V. Ardizzone, M. Pugliese, V. Maiorano, G. Gigli, C. Leblanc, G. Malpuech, D. Solnyshkov, and D. Sanvitto, “Chromodynamics of photons in an artificial non-Abelian magnetic Yang-Mills field”, *arXiv:1912.09684 [cond-mat],* 2019.

[39] S. Donati, L. Dominici, G. Dagvadorj, D. Ballarini, M. De Giorgi, A. Bramati, G. Gigli, Y. G. Rubo,

M. H. Szymanska, and D. Sanvitto, “Twist of generalized skyrmions and spin vortices in a polariton superfluid”, *Proc. Natl. Acad. Sci.*, vol. 113, no. 52, 2016.

[40] O. Bleu, D. D. Solnyshkov, and G. Malpuech, “Measuring the quantum geometric tensor in two-dimensional photonic and exciton-polariton systems”, *Phys. Rev. B*, vol. 97, 2018.

[41] A. Kavokin, J. J. Baumberg, G. Malpuech, and F. P. Laussy, “Microcavities”, *Oxford University Press*, 2011.

[42] J. Ren, Q. Liao, F. Li, Y. Li, O. Bleu, G. Malpuech, J. Yao, H. Fu, and D. Solnyshkov, “Nontrivial band geometry in an optically active system”, *arXiv:1912.05994 [cond-mat]*, 2019.

[43] S. Waltersperger, V. Olivier, C. Pradervand, W. Glettig, M. Salathe, M. R. Fuchs, A. Curtin, X. Wang, S. Ebner, E. Panepucci, T. Weinert, C. Schulze-Briese, M. Wang, “PRIGo: a new multi-axis goniometer for macromolecular crystallography”, *J. Synchrotron Rad.*, vol. 22, no. 4, 2015.

[44] W. Kabsch, “*XDS*”, *Acta Cryst. D*, vol. 66, no.2, 2010.

[45] C. Giacovazzo, Phasing in Crystallography: A Modern Perspective, *Oxford University Press*, 2014.

[46] M. C. Burla, R. Caliandro, B. Carrozzini, G. L. Cascarano, C. Cuocci, C. Giacovazzo, M. Mallamo, A. Mazzone, G. Polidori, “Crystal structure determination and refinement *via SIR2014*”, *J. Appl. Cryst.*, vol. 48, no. 1, 2015.

[47] G. M. Sheldrick, “*SHELXT*- Integrated space-group and crystal-structure determination”, *Acta Cryst. A*, vol. 71, no. 1, 2015.

# Supplementary Material: Tuning the Berry curvature in 2D Perovskite

Laura Polimeno[1,2], Milena De Giorgi[2], Giovanni Lerario[2], Luisa De Marco[2], Lorenzo Dominici[2], Vincenzo Ardizzone[2], Marco Pugliese[1,2], Carmela T. Prontera[2], Vincenzo Maiorano[2], Anna Moliterni[3], Cinzia Giannini[3], Vincent Olieric[4], Giuseppe Gigli[1,2], Dario Ballarini[2], Dmitry Solnyshkov[5,6], Guillaume Malpuech[5], Daniele Sanvitto[2]

[1]Dipartimento di Matematica e Fisica, "Ennio de Giorgi", Università Del Salento, Campus Ecotekne, via Monteroni, Lecce, 73100, Italy.

[2]CNR NANOTEC, Institute of Nanotechnology, Via Monteroni, 73100 Lecce, Italy.

[3]Istituto di Cristallografia, CNR, Via Amendola, 122/O Bari, 70126 Italy.

[4] Paul Scherrer Institute, Forschungstrasse 111, Villigen-PSI, 5232, Switzerland.

[5]Institut Pascal, PHOTON-N2, Université Clermont Auvergne, CNRS, SIGMA Clermont, F-63000 Clermont-Ferrand, France.

[6]Institut Universitaire de France (IUF), 75231 Paris, France.

## I. X-RAY DIFFRACTION

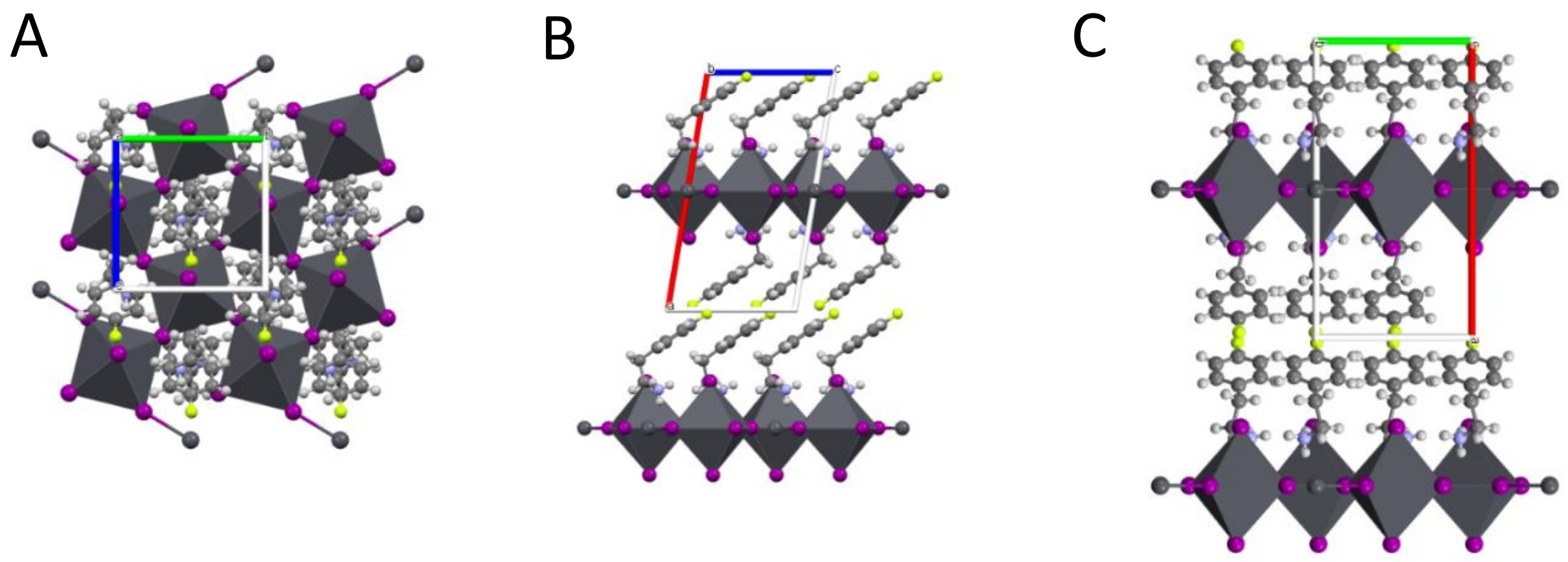

Figure S1: Crystal structure of PEAI-F. A), B) and C): views of the crystal packing along *a* (A), *b* (B) and *c* (C) planes. C, N, F, Pb and I are drawn as grey, light blue, light green, dark grey and purple spheres, respectively, while H as white spheres.

PEAI-F cristallizes in the centrosymmetric space group $P2_1/c$. The crystal packing consists of layers of anionic inorganic networks of corner-sharing $[PbI_6]^{2-}$ octahedra, sandwiched between two layers of fluorophenethylammonium cations (see Figure S1). The layered perovskite is oriented along the longest axis (i.e., the *a* axis); the distance between two successive equatorial inorganic layers is 16.27 Å, and that one between F1-F1 non bonded atoms is 3.415 Å. These data are in accordance with the results obtained by Kikuchi et al. [1]. Main crystallographic data are given in Table S1; additional data concerning refined geometric parameters (bond distances and angles) are provided in Table S2. The interactions between organic cations and inorganic anions and, in particular, the presence of directional hydrogen bonds involving organic cations and halogens, have agreat influence in the perovskite-derivative structures and give rise to a coordinated structural distortion [2].

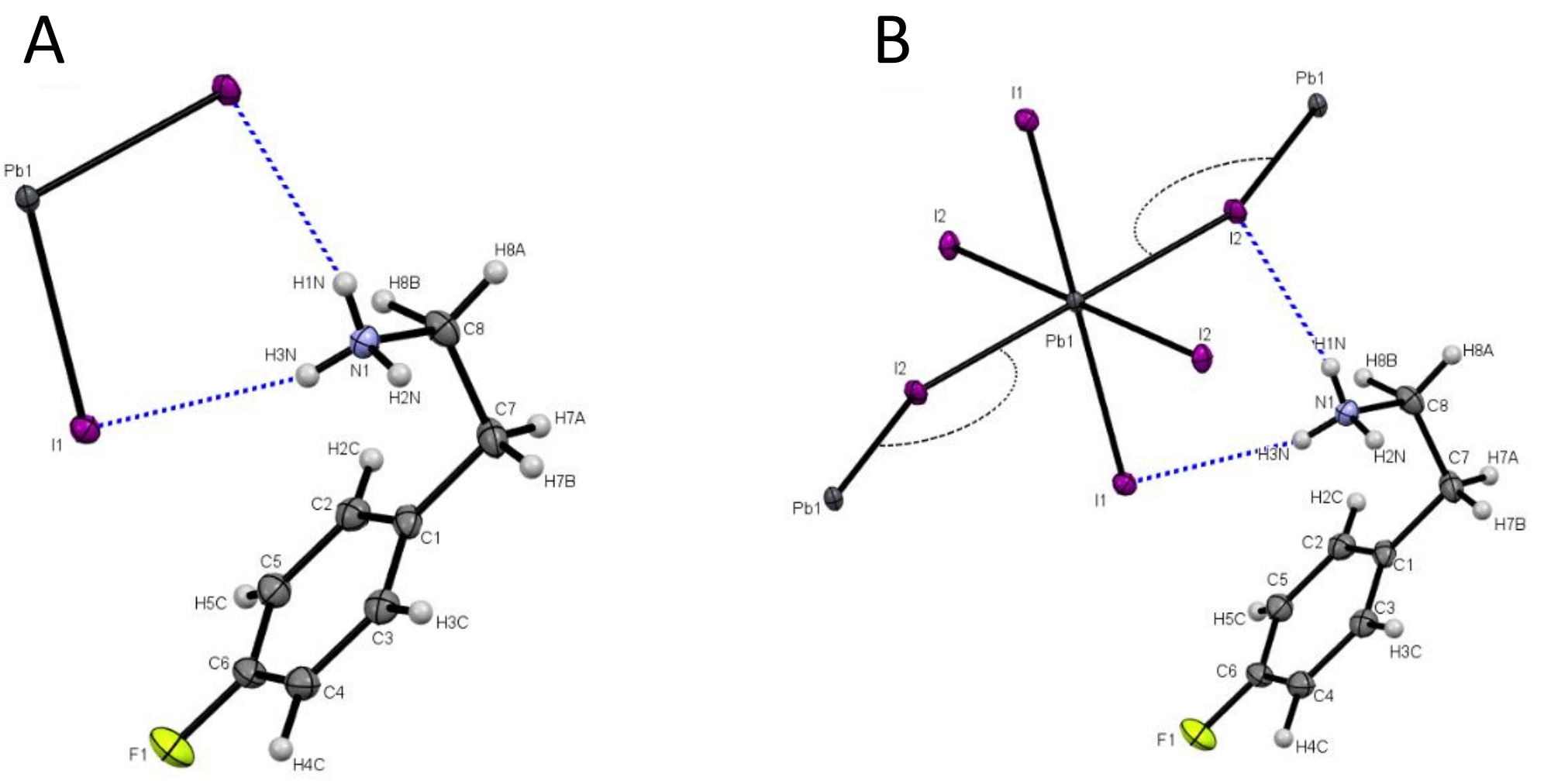

Figure S2: A) A view of the asymmetric unit with the atomic labelling scheme. B) A view of the asymmetric unit plus four symmetry equivalent I atoms to complete the octahedron and two symmetry equivalent Pb atoms to show the distortion of the in-plane inorganic layer [the Pb–I–Pb angle is 151.774(7)°]. The broken blue lines indicate two of the N—H···I hydrogen bonds listed in Table S3. Ellipsoids are drawn at 50% of probability level.

$[PbI_6]^{2-}$ octahedra are in-plane distorted (see Figures S1a and S2b): the in-plane Pb-I-Pb angle is 151.774(7)° (see figure S2b and Table S2), far from the typical value observed in case of undistorted layers (i.e., 180 °), causing an in-plane rotation of the octahedral. [3]

Crystallographic data of PEAI-F have been deposited at the Cambridge Crystallographic Data Centre (CCDC) with deposit number CCDC2013268 and can be obtained free of charge via www.ccdc.cam.ac.uk/structures.

**Table S1:** PEAI-F: main experimental and crystallographic details.

| | |
|---|---|
| *Crystal data* | |
| Chemical formula | $PbI_4 \cdot 2(C_8H_{11}FN)$ |
| $M_r$ | 995.14 |
| Crystal system, space group | Monoclinic, $P2_1/c$ |
| Temperature (K) | 100 |
| *a*, *b*, *c* (Å) | 16.499 (3), 8.555 (2), 8.728 (2) |
| β (°) | 99.50 (2 |
| *V* (Å³) | 1215.1 (5) |
| *Z* | 2 |
| Radiation type | Synchrotron, *λ*=0.72932 Å |
| Crystal size (mm) | 0.09 × 0.05 × 0.05 |
| *μ* ($mm^{-1}$) | 12.76 |
| *Data collection* | |
| Diffractometer | Multi-axis PRIGo goniometer |
| $T_{min}$, $T_{max}$ | 0.246, 1.00 |
| No. of measured, independent and observed [$I > 2\sigma(I)$] reflections | 20197, 3244, 3232 |
| $R_{int}$ | 0.039 |
| $(\sin \theta/\lambda)_{max}$ ($Å^{-1}$) | 0.715 |
| *Refinement* | |
| Bond precision (C–C) (Å) | 0.0054 |
| $R[F^2 > 2\sigma(F^2)]$, $wR(F^2)$, $S$ | 0.030, 0.067, 1.37 |
| No. of reflections | 3244 |
| No. of parameters | 149 |
| H-atom treatment | Only H-atom coordinates refined |
| $\Delta\rho_{max}$, $\Delta\rho_{min}$ (e $Å^{-3}$) | 1.39, −4.36 |

**Table S2** - Additional crystallographic data concerning bond distances and angles

*Geometric parameters (Å, °)*

| | | | |
|---|---|---|---|
| Pb1—I2$^{i}$ | 3.1472 (2) | C4—C6 | 1.376 (5) |
| Pb1—I2$^{ii}$ | 3.1472 (2) | C4—H4C | 1.07 (6) |
| Pb1—I2$^{iii}$ | 3.1538 (2) | C5—C6 | 1.383 (5) |
| Pb1—I2 | 3.1538 (2) | C5—H5C | 0.96 (6) |
| Pb1—I1$^{iii}$ | 3.2132 (2) | C6—F1 | 1.362 (5) |
| Pb1—I1 | 3.2132 (2) | C7—C8 | 1.521 (5) |
| I2—Pb1$^{iv}$ | 3.1472 (2) | C7—H7A | 1.01 (5) |
| C1—C3 | 1.386 (5) | C7—H7B | 0.97 (5) |
| C1—C2 | 1.398 (5) | N1—C8 | 1.495 (5) |
| C1—C7 | 1.514 (6) | N1—H1N | 0.82 (5) |
| C2—C5 | 1.388 (5) | N1—H2N | 0.90 (5) |
| C2—H2C | 0.90 (6) | N1—H3N | 0.84 (5) |
| C3—C4 | 1.397 (5) | C8—H8B | 1.11 (5) |
| C3—H3C | 0.96 (6) | C8—H8A | 0.98 (5) |
| I2$^{i}$—Pb1—I2$^{ii}$ | 180.0 | C6—C4—H4C | 120 (3) |
| I2$^{i}$—Pb1—I2$^{iii}$ | 88.884 (2) | C3—C4—H4C | 122 (3) |
| I2$^{ii}$—Pb1—I2$^{iii}$ | 91.116 (2) | C6—C5—C2 | 118.3 (3) |
| I2$^{i}$—Pb1—I2 | 91.116 (2) | C6—C5—H5C | 118 (3) |
| I2$^{ii}$—Pb1—I2 | 88.884 (2) | C2—C5—H5C | 123 (3) |
| I2$^{iii}$—Pb1—I2 | 180.0 | F1—C6—C4 | 118.8 (3) |
| I2$^{i}$—Pb1—I1$^{iii}$ | 86.973 (5) | F1—C6—C5 | 118.5 (3) |
| I2$^{ii}$—Pb1—I1$^{iii}$ | 93.027 (5) | C4—C6—C5 | 122.7 (4) |
| I2$^{iii}$—Pb1—I1$^{iii}$ | 89.924 (5) | C1—C7—C8 | 112.6 (4) |
| I2—Pb1—I1$^{iii}$ | 90.076 (5) | C1—C7—H7A | 106 (3) |
| I2$^{i}$—Pb1—I1 | 93.027 (5) | C8—C7—H7A | 108 (3) |
| I2$^{ii}$—Pb1—I1 | 86.973 (5) | C1—C7—H7B | 107 (3) |
| I2$^{iii}$—Pb1—I1 | 90.076 (5) | C8—C7—H7B | 109 (3) |
| I2—Pb1—I1 | 89.924 (5) | H7A—C7—H7B | 115 (4) |
| I1$^{iii}$—Pb1—I1 | 180.0 | C8—N1—H1N | 114 (4) |
| Pb1$^{iv}$—I2—Pb1 | 151.774 (7) | C8—N1—H2N | 109 (3) |
| C3—C1—C2 | 119.4 (4) | H1N—N1—H2N | 103 (5) |
| C3—C1—C7 | 120.3 (3) | C8—N1—H3N | 113 (4) |
| C2—C1—C7 | 120.3 (3) | H1N—N1—H3N | 106 (5) |
| C5—C2—C1 | 120.7 (4) | H2N—N1—H3N | 111 (5) |
| C5—C2—H2C | 121 (4) | N1—C8—C7 | 110.7 (3) |
| C1—C2—H2C | 119 (4) | N1—C8—H8B | 99 (3) |
| C1—C3—C4 | 120.7 (3) | C7—C8—H8B | 110 (3) |
| C1—C3—H3C | 119 (3) | N1—C8—H8A | 107 (3) |
| C4—C3—H3C | 121 (3) | C7—C8—H8A | 115 (3) |
| C6—C4—C3 | 118.3 (3) | H8B—C8—H8A | 114 (4) |
| | | | |
| C3—C1—C2—C5 | −0.6 (6) | C3—C4—C6—C5 | 0.9 (6) |
| C7—C1—C2—C5 | −179.4 (4) | C2—C5—C6—F1 | 179.4 (3) |
| C2—C1—C3—C4 | −0.1 (6) | C2—C5—C6—C4 | −1.6 (6) |
| C7—C1—C3—C4 | 178.7 (3) | C3—C1—C7—C8 | −105.6 (4) |
| C1—C3—C4—C6 | 0.0 (5) | C2—C1—C7—C8 | 73.2 (4) |
| C1—C2—C5—C6 | 1.5 (6) | C1—C7—C8—N1 | 58.5 (4) |
| C3—C4—C6—F1 | 179.8 (3) | | |

*Symmetry codes*: (i) *x*, −*y*+1/2, *z*−1/2; (ii) −*x*+1, *y*−1/2, −*z*+3/2; (iii) −*x*+1, −*y*, −*z*+1; (iv) −*x*+1, *y*+1/2, −*z*+3/2.

**Table S3** PEAI-F: Hydrogen-bond geometry (Å,°)

| *D*—H···*A* | *D*—H | H···*A* | *D*···*A* | *D*—H···*A* |
|---|---|---|---|---|
| C4—H4C···F1$^{i}$ | 1.07 (6) | 2.52 (6) | 3.535 (4) | 158 (4) |
| N1—H1N···I2 | 0.82 (5) | 2.98 (5) | 3.592 (3) | 133 (5) |
| N1—H1N···I2$^{ii}$ | 0.82 (5) | 3.20 (5) | 3.650 (3) | 117 (4) |
| N1—H2N···I1$^{iii}$ | 0.90 (5) | 2.73 (5) | 3.626 (3) | 172 (5) |
| N1—H3N···I1 | 0.84 (5) | 2.83 (6) | 3.640 (3) | 162 (4) |
| C8—H8B···I1$^{iv}$ | 1.11 (5) | 3.20 (5) | 4.115 (4) | 140 (4) |
| C8—H8A···I2$^{ii}$ | 0.98 (5) | 3.26 (5) | 3.952 (3) | 129 (4) |

*Symmetry codes*: (i) −*x*+2, *y*−1/2, −*z*+3/2; (ii) −*x*+1, −*y*, −*z*+2; (iii) *x*, −*y*−1/2, *z*+1/2; (iv) *x*, −*y*+1/2, *z*+1/2

## II. PEROVSKITE ABSORPTION AND EMISSION SPECTRA

The photoluminescence and the absorption spectra of a 70 nm thick of perovskite single crystal were recorded at room temperature, using a continuous excitation laser ( $\lambda =$ 488nm) and a white light lamp, respectively.

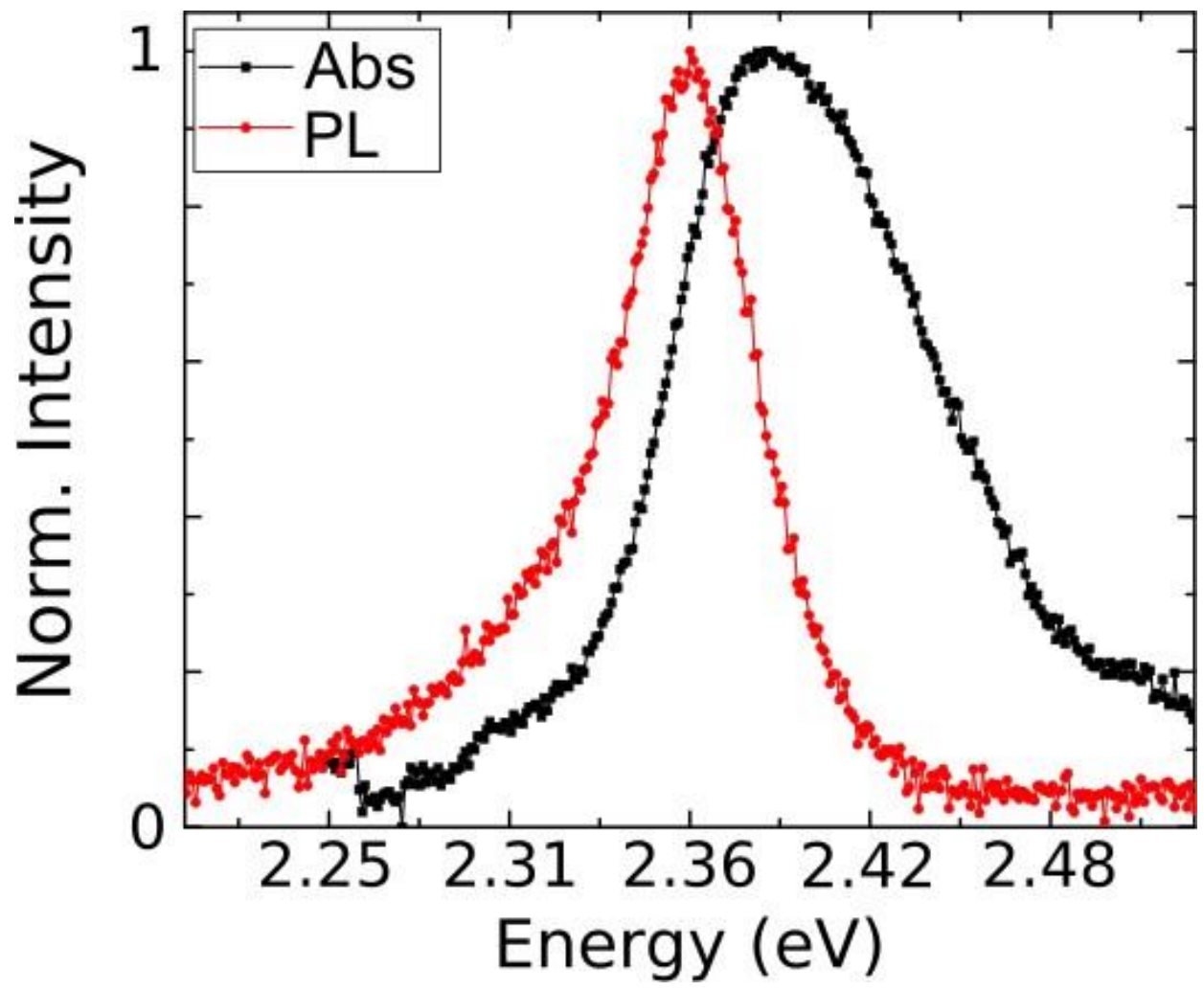

Figure S3: Absorption (black dots) and emission (red dots) spectra of a 70 *nm* thick perovskite single crystal.

## III. ZEEMAN SPLITTING

The values of TE-TM (β) and X-Y (α) energy splitting (Fig. S4B) for different branches was extracted fitting the experimental data reported in Fig. 2A by using the equation (2) of the main text, in absence of the external magnetic field.

After the evaluation of parameters (β) and (α), the Zeeman splitting (Δ) was extracted from the energy gap opening at the Hamiltons' diabolical point when applying 9T (Fig. S4 A).

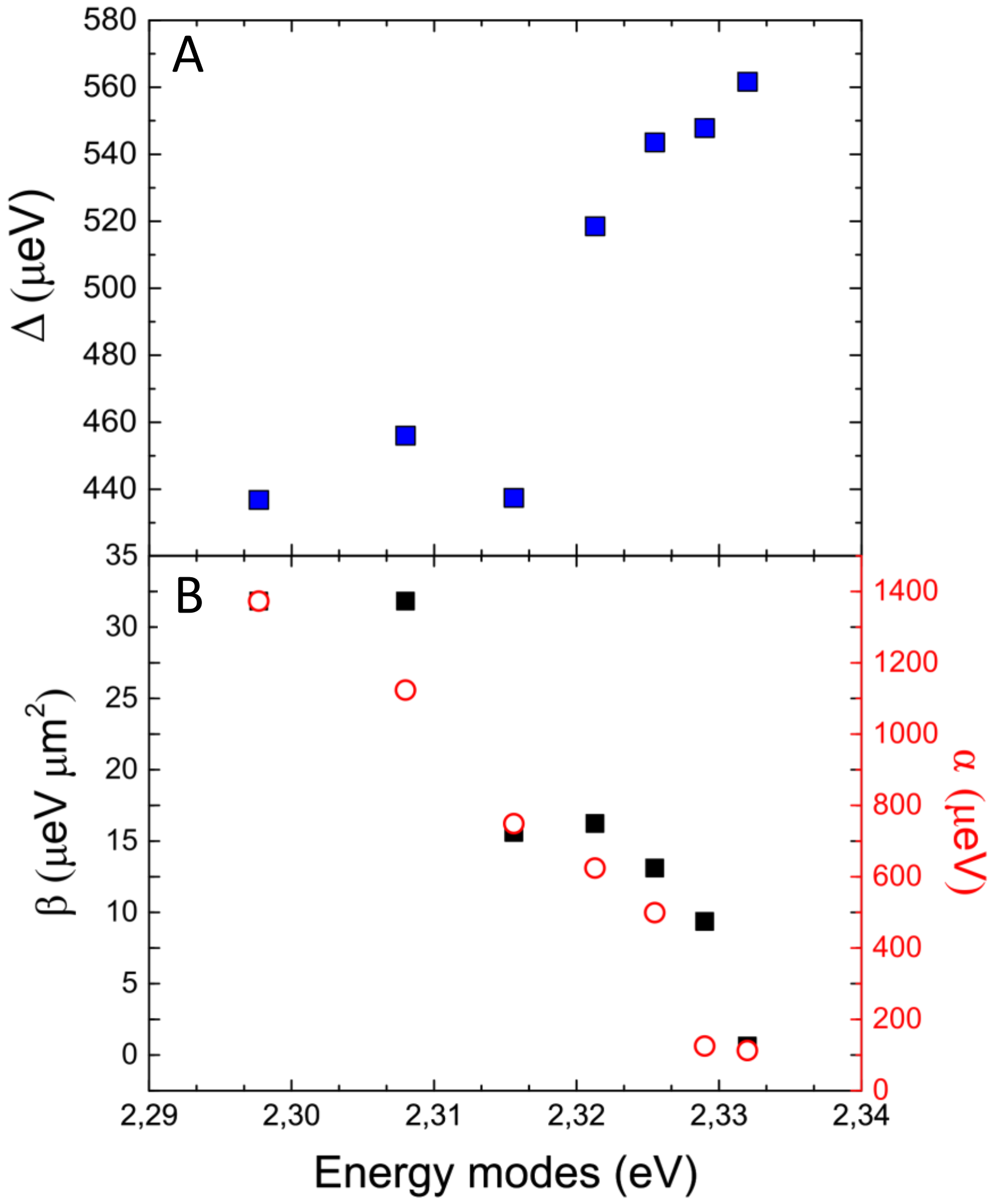

Figure S4: A) Zeeman splitting (Δ) vs polariton energy modes at $B$ = 9 $T$. B) TE-TM ($\beta$, black closed squared points) and X-Y ($\alpha$, red open circular points) splitting for the different polariton energy modes.

## IV. BERRY CURVATURE

In absence of the external magnetic field, the linear polarization precesses along the two crossing rings (Fig. S5 A, B), according to the conical diffraction theory. Fig. S5C shows the experimental Berry curvature extracted from the mode j of the polariton dispersion reported in Fig. S6 at $B = 0\,T$.

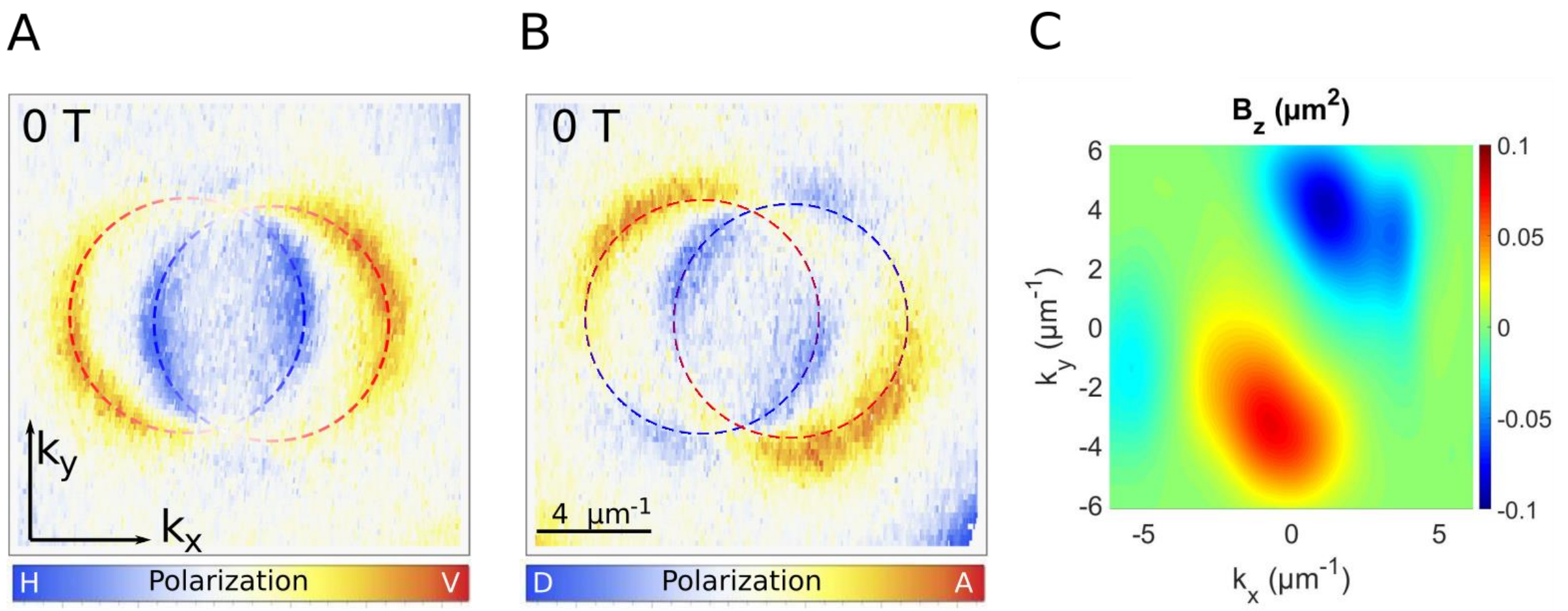

Figure S5: A), B) 2D momentum space mapping of the A) HV and B) DA degree of polarization of the mode 0 ($E$ = 2294 $meV$), reported in the Figure 3 of the main text, in absence of an external magnetic field. C) Experimental Berry curvature extracted from the polarization-resolved measurements for the mode j at $B$ = 0 $T$.

## V. DISPERSION MAP

The optical response of the planar microcavity was investigated in reflection configuration at cryogenic temperature (4 K), under continuous wave excitation (CW 488 nm laser). Multiple lower polariton branches are visible, according to the microcavity free spectral range.

The photoluminescence energy dispersion (Figure S6) shows the lower polariton branches of a 4μm-thick single crystal of PEAI-F.

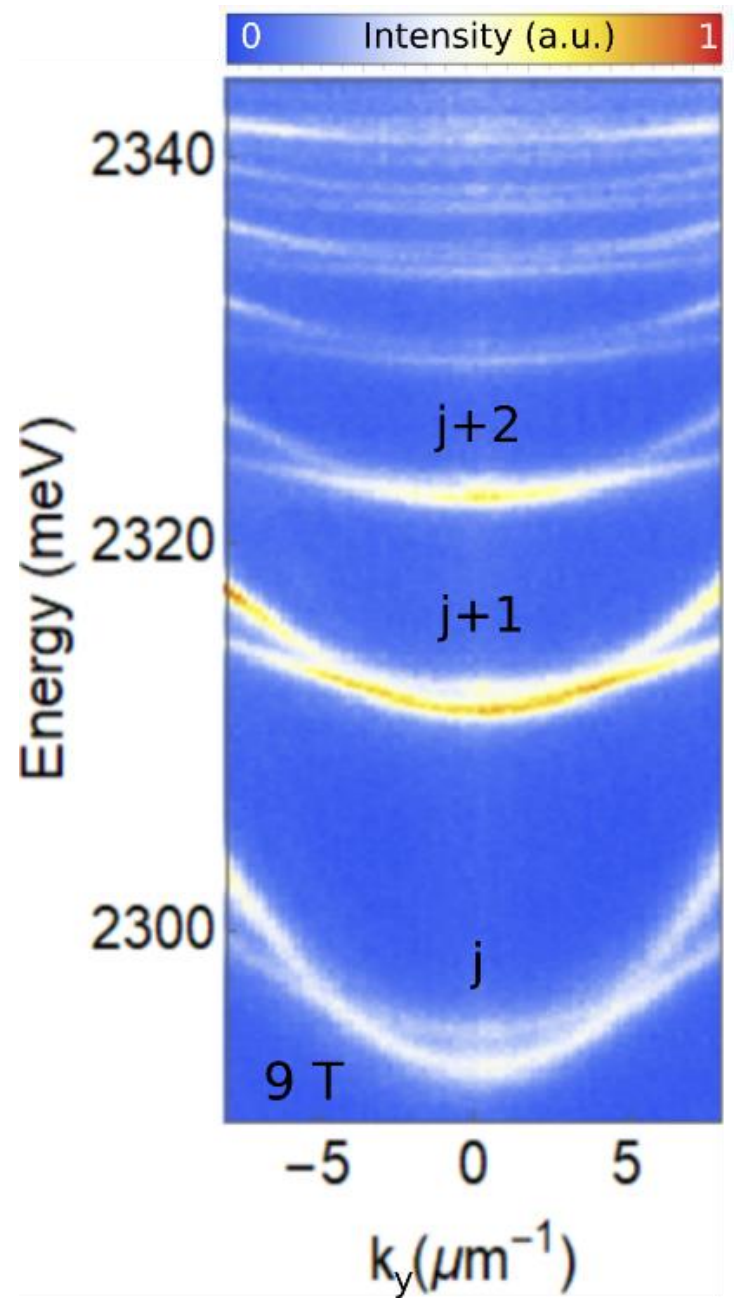

Figure S6: Energy vs the $k_y$ in-plane momentum dispersion maps of the photoluminescence intensity for a 4 $\mu m$-thick perovskite crystal in presence of an external magnetic field, 9 *T.*

**References**

[1] K. Kikuchi, Y. Takeoka, M. Rikukawa, K. Sanui. (2004). *Current Applied Physics*. **4**, 599–602.
[2] Saparov, B., Mitzi, D.B. (2016). *Chem. Rev*. **116**, 4558–4596.
[3] Mercier, N., Louvain, N., Bi, W. (2009). *CrystEngComm*. **11**, 720–734.